\documentclass[showpacs,preprintnumbers,amsmath,onecolumn]{revtex4}

\usepackage{graphicx}
\draft

\begin{document}
\title{Time-Dependent Spintronic Transport and Current-Induced Spin Transfer
Torque in Magnetic Tunnel Junctions}
\author{Zhen-Gang Zhu, Gang Su$^{\ast }$, Qing-Rong Zheng, and Biao Jin}
\address{Department of Physics, The Graduate School of the Chinese Academy of%
\\
Sciences, P.O. Box 3908, Beijing 100039, China}

\begin{abstract}
The responses of the electrical current and the current-induced spin
transfer torque (CISTT) to an ac bias in addition to a dc bias in a magnetic
tunnel junction are investigated by means of the time-dependent
nonquilibrium Green function technique. The time-averaged current
(time-averaged CISTT) is formulated in the form of a summation of dc current
(dc CISTT) multiplied by products of Bessel functions with the energy levels
shifted by $m\hbar \omega _{0}$. The tunneling current can be viewed as to
happen between the photonic sidebands of the two ferromagnets. The electrons
can pass through the barrier easily under high frequencies but difficultly
under low frequencies. The tunnel magnetoresistance almost does not vary
with an ac field. It is found that the spin transfer torque, still being
proportional to the electrical current under an ac bias, can be changed by
varying frequency. Low frequencies could yield a rapid decrease of the spin
transfer torque, while a large ac signal leads to both decrease of the
electrical current and the spin torque. If only an ac bias is present, the
spin transfer torque is sharply enhanced at the particular amplitude and
frequency of the ac bias. A nearly linear relation between such an amplitude
and frequency is observed.
\end{abstract}

\maketitle

\section{Introduction}

The spin-polarized electrical transport in magnetic multilayer structures
has received much attention both experimentally and theoretically in the
last several years\cite{wolf}. The yield of the extensive investigation
promises fascinating implication for applications in information technology.
For instance, people may make use of the different resistive states
corresponding to parallel and antiparallel magnetizations of different
layers as memory elements in magnetic random-access memories (MRAM). It is
now known that the antiparallel alignments of the moments in magnetic layers
lead to a higher electrical resistance than the parallel alignments, giving
rise to a so-called giant magnetoresistance (GMR) or tunnel
magnetoresistance (TMR) depending on different magnetic junction systems.
The origin of these phenomena is widely believed to be mainly caused by the
spin-dependent scatterings of the conduction electrons. As is pointed out
\cite{heide}, when the junction bias is increased, the phenomenon is
significantly reduced. Generally speaking, the resistive state can be
changed by applying an external magnetic field, because the latter could
lead to variations of the traversing paths of the conduction electrons.
However, it is now accepted that not only the electrical current is strongly
affected by a magnetic state, but also the electrical current can conversely
control the magnetic state of the magnetic junctions. This effect is
predicted by Slonczewski \cite{slonczewski} and Berger \cite{berger} in
magnetic multilayer systems by noting that a spin-polarized electrical
current can transfer local spin angular momenta of incident electrons to the
scattering ferromagnet, thereby exerting a torque on the magnetic moments
and therefore changing the magnetic state. When the current is large enough,
the magnetization of ferromagnetic layer can be switched. As a result, the
spin transfer effect may provide a mechanism for a current-controlled
magnetic memory element. Experiments have given evidence of this effect in
the Cu/Co multilayers\cite{tsoi}, nickel nanowires \cite{wegrowe}, manganite
junctions \cite{sun}, point contact magnetic multilayer devices\cite{myers},
bulk manganese oxides\cite{asamitsu}, tunnel junctions\cite{heide}, and
Co/Cu/Co spin valve devices \cite{katine}, etc. A thermally activated
switching of magnetic domains in a Co/Cu/Co spin valve is also confirmed
\cite{myers1}.

To deal with the spin transfer effect, it is useful to introduce the
concepts such as the spin current and the spin torque to describe the
coupling between the conduction electrons and the magnetic moments of
ferromagnetic materials. Those are first proposed by Slonczewski \cite{s2}
based on a quantum-mechanical model. \ Then, this interaction is derived
from the s-d coupling and manifests itself in magnetic multilayers\cite%
{slonczewski} or bulk ferromagnets\cite{zhang}. It also gives a contribution
as a current-induced force on a domain wall or an interaction between spin
waves and itinerant electrons\cite{berger}. The concepts are also used to
deal with a variety of the structures such as the ferromagnet-normal
metal-ferromagnet (FM-NM-FM) junctions \cite{fnf,waintal}, the
ferromagnet-superconductor-ferromagnet junctions \cite{fsf}, and a trilayer
FM-NM-FM contacting a normal metal lead or a superconductor lead\cite%
{waintal1,waintal2}, and so on. Apart from the quantum-mechanical method,
the scattering matrix method is improved to cover this issue\cite%
{waintal,waintal1,waintal2}. Heide {\it et al} derived a set of coupled
Landau-Lifshitz equations for the ferromagnetic layers by considering the
nonequilibrium exchange interaction between layers\cite{heide1}. Zhang {\it %
et al} dealt with the dynamic spin transfer torque and thermally assisted
magnetization reversal by means of the Landau-Lifshitz-Gilbert equation\cite%
{szhang,szhang1} in which a term describing the spin transfer torque is
introduced. The nonequilibrium Green function is also applied to this
question to consider the spin-flip scattering effect on the current-induced
spin transfer torque\cite{zhu1}.

So far, the electrical current, the spin transfer effect and the spin
current in magnetic multilayer systems are extensively studied under an dc
bias voltage, and the investigation under an ac bias is still sparse.
Whether an ac bias voltage applied to the magnetic tunnel junctions could
induce some unusual properties (see e.g.\cite{su}), is still not clear. In
this paper, we shall show the effect of an ac bias on the electrical current
and the spin transfer torque in a magnetic tunnel junction. Since the
switching of magnetic domains depends on the magnitude and the directions of
the spin transfer torque, while the direction of the spin transfer torque is
related to the direction of the electrical current, it is of interest to
anticipate that the response of the system under an ac bias is different
from that under an dc bias. We have found that the spin transfer torque is
sharply enhanced by applying an ac electrical field with a particular
amplitude and frequency.

The rest of this paper is organized as follows. In Sec. II, a model is
proposed, and the necessary formalism under an ac bias is established. In
Sec. III, the tunneling current under an ac electrical field is derived and
studied. In Sec. IV, the current-induced spin transfer torque under an ac
electrical field is investigated. Finally, a brief summary is given in Sec.
V.

\section{Model and Formalism}

\subsubsection{Model}

Let us consider a magnetic tunnel junction in which two ferromagnets (FM)
which are stretched to infinite are separated by a thin insulator (I). The
molecular field in the left ferromagnet is assumed to align along $z$ axis
which is in the junction plane, while the orientation of the molecular field
in the right ferromagnet deviates the $z$ axis by an angle $\theta ,$ which
is along the $z^{\prime }$ axis (such that the frame $x^{\prime }oz^{\prime
} $ deviates $xoz$ by an angle $\theta $). The tunnel current flows along
the $x$ axis which is perpendicular to the junction plane. The Hamiltonian
reads
\begin{equation}
H=H_{L}+H_{R}+H_{T},  \label{Thamiltonian}
\end{equation}%
with
\begin{eqnarray}
H_{L} &=&\sum_{k\sigma }\varepsilon _{k\sigma }^{L}(t)a_{k\sigma }^{\dagger
}a_{k\sigma },  \label{respectH} \\
H_{R} &=&\sum_{q\sigma }[(\varepsilon _{{\bf q}R}{\bf (}t{\bf )-\sigma }%
M_{2}\cos \theta )c_{q\sigma }^{\dagger }c_{q\sigma }-M_{2}\sin \theta
c_{q\sigma }^{\dagger }c_{q\overline{\sigma }}],  \nonumber \\
H_{T} &=&\sum_{kq\sigma \sigma ^{\prime }}[T_{kq}^{\sigma \sigma ^{\prime
}}a_{k\sigma }^{\dagger }c_{q\sigma ^{\prime }}+T_{kq}^{\sigma \sigma
^{\prime }}{}^{\ast }c_{q\sigma ^{\prime }}^{\dagger }a_{k\sigma }],
\nonumber
\end{eqnarray}%
where $a_{k\sigma }$ and $c_{k\sigma }$ are annihilation operators of
conduction electrons with momentum $k$ and spin $\sigma $ $(=\pm 1)$ in the
left and right ferromagnets, respectively. When the time-dependent bias
voltage is applied, the single-particle energies in left and right
ferromagnets, $\varepsilon _{k\sigma }^{L,R}(t)$, become time-dependent\cite%
{haug,jauho}: $\varepsilon _{k\sigma }^{L}(t)=\varepsilon _{{\bf k}L}(t{\bf %
)-\sigma }M_{1},$ $\varepsilon _{{\bf k}L}(t{\bf )=}\varepsilon
_{kL}^{0}+\Delta _{L}(t){\bf -}eV_{0},$ $\Delta _{L}(t)=eV_{L}\cos \omega
_{0}t,$ $M_{1}=\frac{g\mu _{B}h_{L}}{2},$ $\varepsilon _{{\bf q}R}{\bf (}t%
{\bf )=}\varepsilon _{{\bf q}R}^{0}+\Delta _{R}(t),$ $\Delta
_{R}(t)=eV_{R}\cos \omega _{0}t,$ $M_{2}=\frac{g\mu _{B}h_{R}}{2},$ where $%
V_{0}$ is the applied dc bias, $\Delta _{L}(t)$ and $\Delta _{R}(t)$ are
from the applied ac bias, $g$ is the Land\'{e} factor, $\mu _{B}$ is the
Bohr magneton, $h_{L(R)}$ is the magnitude of the molecular field of the
left (right) ferromagnet, $\varepsilon _{{\bf k}L({\bf q}R)}^{0}$ is the
single-particle dispersion of the left (right) FM electrode, $T_{kq}^{\sigma
\sigma ^{\prime }}$ denotes the spin and momentum dependent tunneling
amplitude through the insulating barrier. Note that the spin-flip scattering
is included in $H_{T}$ when $\sigma ^{\prime }=\bar{\sigma}=-\sigma $. It is
this term that violates the spin conservation in the tunneling process.

\subsubsection{Green Functions of Uncoupled leads}

First let us write down the lesser Green function and the retarded
(advanced) Green function for the isolated leads which will be used
subsequently. The single-particle energies of the isolated leads for spin up
and down are splitting, i.e., $\varepsilon _{k\sigma }^{L,0}=\varepsilon
_{kL}-\sigma M_{1}(\sigma =\pm 1,$ corresponding to $\uparrow ,\downarrow ).$
Following the standard procedure, it is not difficult to obtain the lesser
Green function for the left FM lead

\begin{eqnarray}
g_{kL}^{<}(t,t^{\prime }) &=&\left(
\begin{array}{cc}
if_{L}(\varepsilon _{k\uparrow }^{L,0})e^{-i(\varepsilon _{k\uparrow
}^{L,0}-eV_{0})(t-t^{\prime })} & 0 \\
0 & if_{L}(\varepsilon _{k\downarrow }^{L,0})e^{-i(\varepsilon _{k\downarrow
}^{L,0}-eV_{0})(t-t^{\prime })}%
\end{array}%
\right) e^{-i\int_{t^{\prime }}^{t}dt_{1}\Delta _{L}(t_{1})}  \nonumber \\
&=&\sum_{m,n}J_{m}(\frac{eV_{L}}{\omega _{0}})J_{n}(\frac{eV_{L}}{\omega _{0}%
})\left(
\begin{array}{cc}
if_{L}(\varepsilon _{k\uparrow }^{L,0})e^{-i(\varepsilon _{k\uparrow
}^{L,0}-eV_{0})(t-t^{\prime })} & 0 \\
0 & if_{L}(\varepsilon _{k\downarrow }^{L,0})e^{-i(\varepsilon _{k\downarrow
}^{L,0}-eV_{0})(t-t^{\prime })}%
\end{array}%
\right) \cdot  \nonumber \\
&&e^{-i\omega _{0}(mt-nt^{\prime })},
\end{eqnarray}%
where $f_{L}(x)$ is the Fermi function of the left uncoupled lead, $J_{m}(%
\frac{eV_{L}}{\omega _{0}})$ is the $m$-th order of Bessel function.
Throughout this paper, $\hbar =1$ is assumed. Similarly, we can write down
the lesser Green function for the right lead. The retarded and advanced
Green functions have the form of
\begin{eqnarray}
g_{k_{\alpha }}^{r(a)}(t,t^{\prime }) &=&\mp i\theta (\pm t\mp t^{\prime
})\sum_{mn}J_{m}(\frac{eV_{\alpha }}{\omega _{0}})J_{n}(\frac{eV_{\alpha }}{%
\omega _{0}})e^{\pm i\omega _{0}(mt-nt^{\prime })}\cdot  \label{reandadgf} \\
&&\left(
\begin{array}{cc}
e^{\pm i(\varepsilon _{k\uparrow }^{\alpha ,0}-eV_{0})(t-t^{\prime })} & 0
\\
0 & e^{\pm i(\varepsilon _{k\downarrow }^{\alpha ,0}-eV_{0})(t-t^{\prime })}%
\end{array}%
\right) ,  \nonumber
\end{eqnarray}%
where $\alpha =L,R$. A further evaluation will make use of the double time
Fourier transform as follows\cite{anantram,bwang}
\begin{equation}
F(E_{1},E_{2})=\int_{-\infty }^{\infty }dt_{1}\int_{-\infty }^{\infty
}dt_{2}F(t_{1},t_{2})\exp [i(E_{1}t_{1}-E_{2}t_{2})].  \label{dfourier}
\end{equation}%
The above equations will be invoked in the following analysis.

\section{Tunneling Electrical Current Under an ac Bias}

In this section, we shall investigate the tunneling electrical current in
FM-I-FM junctions under an ac electrical field. Following the method in
Refs. \cite{haug,jauho}, the tunneling current can be expressed as
\begin{equation}
I(t)=\frac{2e}{\hbar }%
%TCIMACRO{\func{Re}}%
%BeginExpansion
\mathop{\rm Re}%
%EndExpansion
\sum_{kq}Tr_{\sigma }[{\bf \Omega G}_{kq}^{<}(t,t)],  \label{current}
\end{equation}%
where ${\bf \Omega }={\bf TR,}$ ${\bf T}=\left(
\begin{array}{cc}
T_{1} & T_{2} \\
T_{3} & T_{4}%
\end{array}%
\right) $ with the elements $T_{i}$ ($i=1,...,4$) of the tunneling matrix
which are assumed to be independent of $k$ and $q$ for the sake of simplicity%
\cite{note1}, which is reasonable in the assumption of a wide-band limit
(WBL)\cite{haug,jauho,tkng}, and the elements $T_{2}$ and $T_{3}$ describe
the effect of spin-flip scatterings\cite{zhu}, ${\bf R}=\left(
\begin{array}{cc}
\cos \frac{\theta }{2} & -\sin \frac{\theta }{2} \\
\sin \frac{\theta }{2} & \cos \frac{\theta }{2}%
\end{array}%
\right) ,$ $Tr_{\sigma }$ stands for the trace of the matrix taking over the
spin space, and ${\bf G}_{kq}^{<}(t,t^{\prime })$ is the lesser Green
function in spin space defined as

\begin{equation}
{\bf G}_{kq}^{<}(t,t^{\prime })=\left(
\begin{array}{cc}
G_{kq}^{\uparrow \uparrow ,<}(t,t^{\prime }) & G_{kq}^{\downarrow \uparrow
,<}(t,t^{\prime }) \\
G_{kq}^{\uparrow \downarrow ,<}(t,t^{\prime }) & G_{kq}^{\downarrow
\downarrow ,<}(t,t^{\prime })%
\end{array}%
\right) ,  \label{lesserGF}
\end{equation}%
where $G_{kq}^{\sigma \sigma ^{\prime },<}(t,t^{\prime })\equiv i\langle
c_{q\sigma }^{\dagger }(t^{\prime })a_{k\sigma ^{\prime }}(t)\rangle .$ By
using the nonequilibrium Green function technique \cite{haug,jauho}, we can
get ${\bf G}_{kq}^{<}(t,t^{\prime })$ to the first order:
\begin{equation}
{\bf G}_{kq}^{<}(t,t^{\prime })=\int dt_{1}[g_{qR}^{r}(t,t_{1}){\bf \Omega }%
^{\dagger }g_{kL}^{<}(t_{1},t^{\prime })+g_{qR}^{<}(t,t_{1}){\bf \Omega }%
^{\dagger }g_{kL}^{a}(t_{1},t^{\prime })],  \label{lessergf}
\end{equation}%
where $g_{kL}^{r},$ $g_{qR}^{a},$ $g_{kL}^{<},$ and $g_{qR}^{<}$ are
retarded, advanced, lesser Green function of the left and the right
ferromagnets, respectively. Then, substituting (\ref{lessergf}) into (\ref%
{current}), and making the double-time Fourier transform of Green functions,
we have
\begin{equation}
I(t)=\frac{\pi e}{\hbar }\sum_{mnm^{\prime }n^{\prime }}\int
dEJ(_{mn}^{m^{\prime }n^{\prime }})[f_{L}(E^{\prime \prime
})-f_{R}(E^{\prime })]\Lambda (E^{\prime },E^{\prime \prime },\theta )\cos
(l\omega _{0}t),  \label{timec}
\end{equation}%
where $J(_{mn}^{m^{\prime }n^{\prime }})=J_{m}(\frac{eV_{L}}{\omega _{0}}%
)J_{n}(\frac{eV_{L}}{\omega _{0}})J_{m^{\prime }}(\frac{eV_{R}}{\omega _{0}}%
)J_{n^{\prime }}(\frac{eV_{R}}{\omega _{0}}),$ $E^{\prime \prime }=E-m\omega
_{0}+eV_{0},$ $E^{\prime }=E-n^{\prime }\omega _{0},$ $l=m+m^{\prime
}-n-n^{\prime },$ $\Lambda (E^{\prime },E^{\prime \prime },\theta
)=Tr_{\sigma }[{\bf \Omega D}_{R}(E^{\prime }){\bf \Omega }^{\dagger }{\bf D}%
_{L}(E^{\prime \prime })],$ ${\bf D}_{L(R)}=\left(
\begin{array}{cc}
D_{L(R)\uparrow } & 0 \\
0 & D_{L(R)\downarrow }%
\end{array}%
\right) $ with $D_{L(R)\uparrow (\downarrow )}(\varepsilon
)=D_{L(R)}(\varepsilon \pm M_{1(2)})$ the density of states (DOS) of the
conduction electrons with spin up and down in the left (right) ferromagnet.
In Eq. (\ref{timec}), the real part of the current is taken. The time
average of the electrical current $I(t)$ can be defined as $\left\langle
I(t)\right\rangle =\frac{1}{T_{0}}\int_{-\frac{T_{0}}{2}}^{\frac{T_{0}}{2}%
}dtI_{L}(t)$ with $T_{0}$ the time interval between which the physical
quantity is measured. Thus, the time-averaged tunneling current is given by

\begin{equation}
I_{ac}^{aver}=\left\langle I(t)\right\rangle =\left\{
\begin{array}{clcc}
\sum_{mnm^{\prime }n^{\prime }}I(_{m^{\prime }n^{\prime
}}^{mn})=\sum_{mnm^{\prime }n^{\prime }}J(_{mn}^{m^{\prime }n^{\prime
}})I_{dc}(V_{0}-m\omega _{0}/e,\text{ }n^{\prime }\omega _{0}/e), & l=0; &
&  \\
0, & l\neq 0; &  &
\end{array}%
\right.  \label{averagecu}
\end{equation}%
where $I(_{m^{\prime }n^{\prime }}^{mn})=J(_{mn}^{m^{\prime }n^{\prime
}})I_{dc}(V_{0}-m\omega _{0}/e,$ $n^{\prime }\omega _{0}/e),$ $%
I_{dc}(eV_{0}-m\omega _{0},n^{\prime }\omega _{0})=\frac{\pi e}{\hbar }\int
dE[f_{L}(E^{\prime \prime })-f_{R}(E^{\prime })]\Gamma _{I}(E^{\prime
},E^{\prime \prime })\{1$ $+P_{2}(E^{\prime })[P_{1}(E^{\prime \prime })\cos
\theta +P_{3}(E^{\prime \prime })\sin \theta ]\},$ $P_{1}=\frac{D_{L\uparrow
}(T_{1}^{2}-T_{2}^{2})-D_{L\downarrow }(T_{4}^{2}-T_{3}^{2})}{D_{L\uparrow
}(T_{1}^{2}+T_{2}^{2})+D_{L\downarrow }(T_{3}^{2}+T_{4}^{2})},$ $P_{2}=\frac{%
D_{R\uparrow }-D_{R\downarrow }}{D_{R\uparrow }+D_{R\downarrow }},$ $P_{3}=%
\frac{2(D_{L\uparrow }T_{1}T_{2}+D_{L\downarrow }T_{3}T_{4})}{D_{L\uparrow
}(T_{1}^{2}+T_{2}^{2})+D_{L\downarrow }(T_{3}^{2}+T_{4}^{2})},$ and $\Gamma
_{I}(E^{\prime },E^{\prime \prime })=[D_{R\uparrow }(E^{\prime
})+D_{R\downarrow }(E^{\prime })][D_{L\uparrow }(E^{\prime \prime
})(T_{1}^{2}+T_{2}^{2})+D_{L\downarrow }(E^{\prime \prime
})(T_{3}^{2}+T_{4}^{2})].$ It can be seen that the average ac tunneling
current in a magnetic tunnel junction is modulated via Bessel functions.
This result is quite similar to Eq. (3.3) in Ref.\cite{Tucker} in which the
superconductor tunnel junctions are investigated. By analogy, the time
dependence of the wave function for every electron state is modulated by $%
J_{n}(\frac{eV_{L(R)}}{\omega _{0}})e^{\pm in\omega _{0}t}$. Each
single-particle level is modulated in terms of a probability $J_{n}(\frac{%
eV_{L(R)}}{\omega _{0}})$ and is displaced in energy by $n\hbar \omega _{0}.$
These displacements in energy contributing to the amplitude of the average
tunneling current are equivalent to that dc voltages $(V_{0}-m\omega _{0}/e)$
applied across the junction with a probability $J(_{mn}^{m^{\prime
}n^{\prime }}).$ In Eq.(\ref{averagecu}), an explicit relation between the
time averaged ac tunneling current and the dc current is given, where $l=0$
gives a nontrivial result. This result implies that only the elastic
transmission through the tunnel barrier contributes to the average current,
and the net number of photons absorbed from the ac field must be zero\cite%
{jauho1}. So, every term in the summation of Eq.(\ref{averagecu}) describes
the tunneling process that the electrons tunnel from the excited states $%
eV_{0}-m\omega $ absorbed $m$ photons of the left ferromagnet to the excited
states $n^{\prime }\omega _{0}$ absorbed $n^{\prime }$ photons of the right
one. In this way, the summation is taken over all the excited states of the
left and the right ferromagnets. The ac field gives a correction to the
transition rate $\Gamma _{I}$ by adding the product of Bessel functions,
i.e. $J(_{mn}^{m^{\prime }n^{\prime }})$, which describes the probability of
electron population in the excited states. The tunneling current can be
viewed as to happen between the photonic sidebands of the two ferromagnets.

Next, let us present the numerical results of the time-dependent electric
current. Before going on, we shall first give some presumptions. A parabolic
dispersion for band electrons is assumed, on which is based that the DOS of
conduction electrons are calculated. The Fermi energy and the molecular
field will be taken as $E_{f}=1.295$ eV and $\left\vert {\bf h}%
_{1}\right\vert =\left\vert {\bf h}_{2}\right\vert =0.90$ eV, which are
given in Ref.\cite{moodera} for Fe. We note that the elements of the
coupling matrix $T_{2}$ and $T_{3}$ mean the strength of spin-flip
scatterings which were discussed in Refs. \cite{zhu1,zhu}. In order to focus
our attention on the effect of an ac bias, we shall not consider the effect
of spin-flip scatterings here for brevity and simplicity. So we take $T_{2}$
$=$ $T_{3}=0$, and $T_{1}=T_{4}=0.01eV$. In addition, $V_{R}=0$ is assumed,
then $V_{L}=V_{ac}$. Under this assumption, $I_{ac}^{aver}=\sum_{m}J_{m}^{2}(%
\frac{eV_{ac}}{\omega _{0}})I_{dc}(V_{0}-m\omega _{0}/e,\theta ),$ where $%
J_{m}(\frac{eV_{ac}}{\omega _{0}})$ is the $m$-th Bessel function. When the
ac bias is absent, we get $G_{dc}(V_{0}=0,T=0,\theta )=\partial
I_{dc}/\partial V_{0}=G_{0}[1+$ $P_{2}\sqrt{P_{1}^{2}+P_{3}^{2}}\cos (\theta
-\theta _{f})]$\cite{zhu} at zero dc bias, zero temperature ($T=0$) and $%
\theta $, where $P_{1},P_{2},$ and $P_{3}$ are similar to the formulas as
given before, tan$\theta _{f}=P_{3}/P_{1}$ and $G_{0}=\frac{\pi e^{2}}{%
2\hbar }[(T_{1}^{2}+T_{2}^{2})D_{L\uparrow
}+(T_{3}^{2}+T_{4}^{2})D_{L\downarrow }](D_{R\uparrow }+D_{R\downarrow })$.
As a consequence, $I_{dc}(T=0,\theta )=$ $G_{dc}(0,0,\theta )\cdot V_{0}$
can be used as a scale to measure the ac current.

The time evolution of the tunnel electrical current $I(t)$ in response to an
ac field is depicted in Fig. 1, where $V_{0}=-0.1V$ and $\omega _{0}=0.003$ $%
eV$. As can be seen, when the dc bias is positive, the current can be
negative. When the ac signal is small (e.g. $V_{ac}=0.001V$), the current
varies with time in a cosine manner, which appears to be proportional to the
ac bias. In this case, the current response is similar to the dc case, i.e.
larger the bias, larger the current.\ However, when the ac signal is
stronger, things become different. There appear some resonant peaks, which
can be regarded as to be resulted from the photon-assisted tunneling, and
the tunnel current is still periodic with time. The inset of Fig.1 shows the
case at $V_{ac}=0.01V$ but with a lower frequency for a comparison. It is
found that the peaks split into several peaks in one period of time,
suggesting that the external frequency can change the oscillating frequency
of the tunnel current.

The frequency dependence of the averaged tunneling current is shown in Fig.
2. Small oscillations of the current with frequency can be seen, which are
caused by the summation of the $m$-th current $I(_{m^{\prime }n^{\prime
}}^{mn})$ with different $m$, $n$, $m^{\prime }$, and $n^{\prime }$ owing to
$I(_{m^{\prime }n^{\prime }}^{mn})$ oscillating with the frequency and
having many peaks. The peaks of the $m$-th current correspond to the troughs
of the $(m+1)$-th current. Furthermore, when the ac signal is small (e.g. $%
V_{ac}=0.01V$), the current first increases with small oscillations, and
then, the current almost does not vary with the frequency like a pure
resistance because the pure resistance should not vary with the frequency in
the common sense. However, it can be observed that a large ac signal leads
to a small averaged current also as those expressed in Fig. 3. The reason
for this feature is that an ac bias is imposed on an dc bias, while the
summation current decreases with increasing ac bias. A large $V_{ac}$ means
a large argument of Bessel function, thus leading to a strong modulation to
the current, which makes the current approaching to zero. If $V_{ac}$ is
fixed, a larger $\omega _{0}$ gives a larger averaged current, suggesting
that the current flows easily in this system under a higher frequency. This
character is consistent with the classical feature of systems that metallic
leads are separated by an insulator. Here, we can consider two limiting
cases. When $\omega _{0}\rightarrow 0$, the argument of the Bessel function $%
x_{B}=eV_{ac}/\omega _{0}\rightarrow \infty $, then $J_{m}^{2}(x_{B})\sim
\frac{2}{\pi x_{B}}\cos ^{2}(x_{B}-\frac{\pi }{2}m-\frac{\pi }{4}%
)\rightarrow 0$, which suggests that the current cannot pass easily through
the system. When $V_{ac}\rightarrow 0$ and the frequency becomes larger, the
argument of the Bessel function $x_{B}\rightarrow 0$, then $J_{m\neq
0}(x_{B}\rightarrow 0)\sim 0$ and $J_{0}(0)\sim 1$, which suggests that the $%
m=0$ term is dominant. This case shows that the current can pass
easily through the barrier under higher frequencies. It is the
character of a capacitance in the usual sense.
%$\thicksim $

Let us define $TMR=(J_{P}-J_{AP})/J_{AP}$ by using the real part
of the current. We have investigated the response of $TMR$ to an
ac bias. It is found that TMR almost does not alter with $V_{ac}$
and frequency $\omega _{0} $, with the varying range about in
$0.01\%$ $\sim$ $0.1\%$. To understand this result, we would like
to remark that $TMR$ is mainly contributed by the spin-dependent
scatterings. When electrons from one ferromagnet enter into
another whose magnetization deviates an angle to the first one,
spin up and down electrons bear different potentials. While the ac
field provides the same modulation of quasiparticle levels of the
spin up and spin down electrons shifted by $n\hbar \omega _{0}$,
leading to a change of the magnitude of the current, it affects
less spin-dependent scatterings. If we impose a time-dependent
magnetic field on both electrodes, it is conceivable that $TMR$
would be remarkably influenced by such an ac magnetic field. Work
towards this direction is now in progress.

\section{Current-Induced Spin Transfer Torque Under an ac Bias}

As is known, the relative orientation of magnetizations on both electrodes
can affect considerably the magnitude of the electrical current flowing
through the magnetic tunnel junctions, and meanwhile, the spin-polarized
electrical current can also switch the direction of magnetization of
electrodes. This latter effect comes from an indirect interaction between
ferromagnets, which is caused by the so-coined current-induced spin-transfer
torque (CISTT)\cite{slonczewski,berger}. The physical meaning of the CISTT
can be understood as follows. When the electrons in the first ferromagnet
tunnel through the barrier and enter into the right ferromagnet, the
incoming polarized electrons will precess, eventually to align with the
magnetization direction at an angle $\theta $ in the second ferromagnet. In
this process, there should be a difference of spin angular momenta between
the incoming and outgoing spins in the second ferromagnet. The missing spin
angular momentum must be absorbed by the local moments, thereby generating a
torque which exerts on the moments of the second ferromagnetic layer. This
kind of torque reflects actually the spin angular momentum transfer from the
first ferromagnet to the second by the polarized current. In Ref. \cite{zhu1}%
, we have discussed the CISTT by means of the nonequilibrium Green function
technique under an dc bias. By generalizing our treatment, we now consider
the case under a simultaneous application of both ac and dc biases. The
CISTT is related not only to the available energy levels of both
ferromagnets, but also to the direction of the current. In the ac case, the
available energy levels of ferromagnets will be shifted by absorbing or
emitting photons. The tunneling can be viewed as to occur between the
photonic sidebands of the left and the right ferromagnets, and the magnitude
of the current can be tuned. The ac field can change the current directions
by frequency $2\omega _{0}$. It may be expected that the ac field may impose
considerable effect on the CISTT. The total spin of the right ferromagnet
can be expressed as\cite{zhu1}
\begin{equation}
{\bf s}(t)=\frac{\hbar }{2}%
%TCIMACRO{\underset{k\mu \nu }{\sum }}%
%BeginExpansion
\mathrel{\mathop{\sum }\limits_{k\mu \nu }}%
%EndExpansion
c_{k\mu }^{\dagger }c_{k\nu }({\bf R}^{-1}\chi _{\mu })^{\dagger }\stackrel{%
\wedge }{{\bf \sigma }}({\bf R}^{-1}\chi _{\nu }),  \label{totalsp}
\end{equation}%
where $\stackrel{\wedge }{{\bf \sigma }}$ is the Pauli matrices and $\chi
_{\mu (\nu )}$ is spin states, which is written down in the $xyz$ coordinate
frame while the spins ${\bf s}_{2}$ are quantized in the $x^{\prime
}y^{\prime }z^{\prime }$ frame, and the rotation matrix ${\bf R}$ is the
same as before. Since
\begin{equation}
\stackrel{\cdot }{{\bf s}}_{L,R}\sim I_{e}\widehat{s}_{L,R}\times (\widehat{s%
}_{L}\times \widehat{s}_{R}),  \label{torcurrent}
\end{equation}%
where $\widehat{s}_{L,R}$ are unit vectors $\widehat{s}_{i}=\overrightarrow{%
{\bf s}_{i}}/s_{i}$ and $\hbar \overrightarrow{{\bf s}_{i}}$ represents the
respective total spin momenta per unit area of the ferromagnets\cite%
{slonczewski}, we know that the direction of the spin transfer torque $%
\stackrel{\cdot }{{\bf s}}_{R}$ is just along the $x^{\prime }$ direction in
the $x^{\prime }y^{\prime }z^{\prime }$ coordinate frame. From Eq. (\ref%
{totalsp}) we obtain ${\bf s}_{2}(t)=\frac{\hbar }{2}%
%TCIMACRO{\underset{k\sigma }{\sum }}%
%BeginExpansion
\mathrel{\mathop{\sum }\limits_{k\sigma }}%
%EndExpansion
(c_{k\sigma }^{\dagger }c_{k\overline{\sigma }}\cos \theta -\sigma
c_{k\sigma }^{\dagger }c_{k\sigma }\sin \theta )=s_{2x^{\prime }0}\cos
\theta -s_{2z^{\prime }0}\sin \theta ,$ where $s_{2x^{\prime }0}$ and $%
s_{2z^{\prime }0}$ are $x^{\prime }$- and $z^{\prime }$-components of the
total spins in the $x^{\prime }y^{\prime }z^{\prime }$ coordinate frame in
which the spins ${\bf s}_{2}$ are quantized. So the CISTT can be obtained%
\cite{zhu1}

\begin{equation}
\tau ^{Rx}(t)=-\cos \theta
%TCIMACRO{\func{Re}}%
%BeginExpansion
\mathop{\rm Re}%
%EndExpansion
%TCIMACRO{\underset{kq}{\sum }}%
%BeginExpansion
\mathrel{\mathop{\sum }\limits_{kq}}%
%EndExpansion
Tr_{\sigma }[{\bf G}_{kq}^{<}(t,t)\stackrel{\wedge }{\sigma }_{1}{\bf T}%
^{\dagger }]+\sin \theta
%TCIMACRO{\func{Re}}%
%BeginExpansion
\mathop{\rm Re}%
%EndExpansion
%TCIMACRO{\underset{kq}{\sum }}%
%BeginExpansion
\mathrel{\mathop{\sum }\limits_{kq}}%
%EndExpansion
Tr_{\sigma }[{\bf G}_{kq}^{<}(t,t)\stackrel{\wedge }{\sigma }_{3}{\bf T}%
^{\dagger }],  \label{torquex}
\end{equation}%
where $\stackrel{\wedge }{\sigma }_{1}=\left(
\begin{array}{cc}
0 & 1 \\
1 & 0%
\end{array}%
\right) ,$ $\stackrel{\wedge }{\sigma }_{3}=\left(
\begin{array}{cc}
1 & 0 \\
0 & -1%
\end{array}%
\right) $ are Pauli matrices, and ${\bf G}_{kq}^{<}(t,t^{\prime })$ is the
lesser Green function in spin space defined as above. By using the
nonequilibrium Green function technique \cite{haug} (a similar framework
employed in Ref.\cite{zhu1}), we can get the torque to the first order of
the Green function
\begin{equation}
{\bf G}_{kq}^{<}(t,t^{\prime })=\int dt_{1}[{\bf g}_{kL}^{r}(t,t_{1}){\bf TRg%
}_{qR}^{<}(t_{1},t^{\prime }){\bf R}^{\dagger }+{\bf g}_{kL}^{<}(t,t_{1})%
{\bf TRg}_{qR}^{a}(t_{1},t^{\prime }){\bf R}^{\dagger }],  \label{qqq}
\end{equation}%
where ${\bf g}_{kL}^{r},${\bf \ }${\bf g}_{qR}^{a},${\bf \ }${\bf g}%
_{kL}^{<},$ and ${\bf g}_{qR}^{<}$ are retarded, advanced, lesser Green
function of the isolated left and the right ferromagnets, respectively.
After some algebra, we obtain the Fourier transform of the CISTT

\begin{eqnarray*}
\tau ^{Rx}(\omega ) &=&2\pi ^{2}\sum_{mnm^{\prime }n^{\prime }}\int
dEJ(_{mn}^{m^{\prime }n^{\prime }})[f_{L}(E^{\prime \prime
})-f_{R}(E^{\prime })]\Gamma (E^{\prime },E^{\prime \prime }) \\
&&\cdot \lbrack P_{1}(E^{\prime \prime })\sin \theta -P_{3}(E^{\prime \prime
})\cos \theta ]\delta (\omega +l\omega _{0}),
\end{eqnarray*}%
where $\Gamma (E^{\prime },E^{\prime \prime })=[D_{R\uparrow }(E^{\prime
})+D_{R\downarrow }(E^{\prime })][D_{L\uparrow }(E^{\prime \prime
})(T_{1}^{2}+T_{2}^{2})+D_{L\downarrow }(E^{\prime \prime
})(T_{3}^{2}+T_{4}^{2})].$ Here we may note that the direction of the spin
torque is related not only to whether the applied dc bias is positive or
negative\cite{slonczewski,tsoi,sun,myers,katine} (namely, the direction of
the electrical current in a steady state), but also to the frequency of the
external ac bias and the Bessel function$^{\prime }$s order. The time
average of CISTT may be defined as $\left\langle \tau ^{Rx}\right\rangle =%
\frac{1}{T_{0}}\int_{-\frac{T_{0}}{2}}^{\frac{T_{0}}{2}}dt\tau ^{Rx}(t)$.
Then, we get
\begin{equation}
\tau _{aver}^{Rx}=\left\langle \tau ^{Rx}\right\rangle =\left\{
\begin{array}{clcc}
\pi \sum_{mnm^{\prime }n^{\prime }}\int dEJ(_{mn}^{m^{\prime }n^{\prime
}})[f_{L}(E^{\prime \prime })-f_{R}(E^{\prime })]\Gamma (E^{\prime
},E^{\prime \prime })P(E^{\prime \prime },\theta ), & l=0; &  &  \\
0, & l\neq 0; &  &
\end{array}%
\right.  \label{acav}
\end{equation}%
where $l=m+m^{\prime }-n-n^{\prime }$, $E^{\prime },$ $E^{\prime \prime },$ $%
J(_{mn}^{m^{\prime }n^{\prime }}),$ $P(E^{\prime \prime },\theta
)=[P_{1}(E^{\prime \prime })\sin \theta -P_{3}(E^{\prime \prime })\cos
\theta ]$, and $\Gamma (E^{\prime },E^{\prime \prime })$ are defined as
above.

The $\theta $ dependence of the spin-transfer torque is similar to that in
Refs. \cite{slonczewski,waintal,waintal1,waintal2}. In a colinear case ($%
\theta =0$ or $\pi $) and without spin-flip scatterings, the CISTT
disappears, even though an ac bias is present. At $\theta =\frac{\pi }{2}$,
the CISTT tends to a maximum in the presence of no-flip of spins\cite{zhu1}.
In this case, $\omega _{0}$, $V_{ac},$ and $V_{0}$ can affect the maximum of
the torques. In the following discussions, we suppose that $T_{1}=T_{4}$, $%
T_{2}=T_{3}=0$, $V_{R}=0$ and $V_{L}=V_{ac}$. We shall use $\tau
_{dc}(T=0,\theta )=$ $G_{dc}^{\tau }(0,0,\theta )\cdot V_{0}$ as a scale,
where $G_{dc}^{\tau }(0,0,\theta )=e\pi T_{1}^{2}[D_{R\uparrow
}(E_{f})+D_{R\downarrow }(E_{f})][D_{L\uparrow }(E_{f})+D_{L\downarrow
}(E_{f})]P_{1}(E_{f})\sin \theta $ with $P_{1}(E_{f})=[D_{L\uparrow
}(E_{f})-D_{L\downarrow }(E_{f})]/[D_{L\uparrow }(E_{f})+D_{L\downarrow
}(E_{f})].$

First, in order to observe the effect of an ac bias on the CISTT, we present
$V_{ac}$ dependence of the time-dependent spin-transfer torque at different
frequency $\omega _{0}$ in Fig. 4. It is seen that the amplitude and
oscillating frequency of the torque are modulated. As the direction of the
spin-transfer torque is related to the direction of the electrical current,
it is changed continuously with the ac current. Note that in Fig. 4, an dc
voltage is applied simultaneously together with the ac bias. When $V_{ac}$
is small, the oscillating amplitude becomes large, and when $V_{ac}$ becomes
larger, the torque is strongly suppressed. Besides, one may see that a
larger $\omega _{0}$ gives a small oscillating frequency of the torque with $%
V_{ac}$. The frequency ($\omega _{0}$) dependence of the time-dependent
CISTT at different amplitudes of the ac bias is shown in Fig. 5. It is seen
that the torque oscillates with the external frequency, and it appears that
the signal of the CISTT is strong in some regime of $\omega _{0}$, and is
almost vanishing in other regimes. This feature becomes more evident at a
larger $V_{ac}$.

The time-averaged spin torque $\left\langle \tau ^{Rx}\right\rangle $ is
more interesting. In Fig. 6, the $V_{ac}$ dependence of the time-averaged
CISTT is plotted under different frequencies of the ac bias. One may find
that the time-averaged CISTT increases slowly with small $V_{ac}$, and then
decreases rapidly with small oscillations. The CISTT approaches to zero when
$V_{ac}$ becomes larger. As shown above, it seems that a large amplitude of
the ac bias field may suppress the spin-transfer torque. When $V_{ac}$ is
decreasing, $\left\langle \tau ^{Rx}\right\rangle /\tau _{dc}(0)$ approaches
asymptotically to a single curve for different frequency $\omega _{0}$. At a
given $\omega _{0}$, $\left\langle \tau ^{Rx}\right\rangle /\tau _{dc}(0)$
has a maximum at a specific $V_{ac}$. In this case, the time-averaged CISTT,
$\left\langle \tau ^{Rx}\right\rangle $ (note that we use $\tau _{aver}^{Rx}$
to denote it in figures herefater), in the simultaneous presence of an ac
bias and a dc bias, can be expressed by that under only a dc bias:
\begin{equation}
\left\langle \tau ^{Rx}\right\rangle =\sum_{mnm^{\prime }n^{\prime
}}J(_{mn}^{m^{\prime }n^{\prime }})\tau _{dc}^{Rx}(V_{0}-n^{\prime }\omega
_{0}/e,m\omega _{0}/e),  \label{atoracdc}
\end{equation}%
where the subscript $dc$ represents the quantity under only a dc bias. It is
clear that the Bessel function modulates the amplitude of the torque. From
Eq. (\ref{torcurrent}), one may see that a large dc current generates a
large torque. In the case under an ac bias, the electrical current might
have the similar character. Since the tunneling is viewed as to take place
between the modulated levels of ferromagnets, the modulations lead to
different transmissions, thus enabling the spin-transfer torque to exhibit
various features. This character also manifests itself in Fig. 7, in which
the time-averaged spin-transfer torque as a function of the external
frequency $\omega _{0}$ for different $V_{ac}$ is shown. It can be found
that small frequencies affect the time-averaged CISTT dramatically,
manifested explicitly by an approximately linear relation between the CISTT
and small $\omega _{0}$. In comparison to Fig. 2, it may be concluded that $%
\left\langle \tau ^{Rx}(t)\right\rangle $ is proportional to the
time-averaged electrical current, say $\left\langle I_{e}(t)\right\rangle $,
and the CISTT can be still induced by the ac electrical current. Note that
Eq. (\ref{torcurrent}) was proposed under an dc bias\cite{slonczewski}.
Under a small ac signal, the torque almost does not vary in higher
frequencies, which is saturated with the magnitude larger than that under
only a dc bias.

In above, we have applied simultaneously both the dc and ac biases to the
tunnel junction. In the presence of only an ac bias, the time-averaged CISTT
as a function of $V_{ac}$ is given in Fig. 8. It is seen that the
time-averaged spin transfer torque first increases sharply and then
decreases rapidly with increasing $V_{ac}$. In other words, the CISTT takes
its larger values in a narrow regime of $V_{ac}$. The higher the external
frequency $\omega _{0}$, the larger the maximum of the torque. This is
because the spin transfer torque is related to the subtraction of two Fermi
functions, and the latter is very small when only an ac bias is present. The
principal contribution to the result comes from a narrow regime of $V_{ac}$.
The inset of Fig. 8 gives a plot of $V_{ac}^{p}$ versus $\omega _{0},$ where
$V_{ac}^{p}$ is the specific value of the amplitude of the ac bias at which
the peaks of the torque appear. It is to note that $V_{ac}^{p}$ is almost
proportional to the frequency $\omega _{0}$. Compared to Fig. 6, where the
dc and ac biases are present simultaneously, the peaks in Fig. 8 are more
sharp. These peaks can be regarded as the evidence of the photon-assisted
enhancements of the CISTT. In Fig. 6, the averaged spin transfer torque
increases slightly with increasing $V_{ac}$, and tends to a maximum, then
decreasing rapidly. If the dc bias disappear slowly, the spin transfer
torque tends to zero when $V_{ac}$ tends to zero, while the situation in
Fig. 6 is not. This reveals that the CISTT exhibits different behaviors
under different biases. Our results show that an ac bias could enhance the
CISTT at particular values of the amplitude and frequency.

\section{Discussion and Summary}

We have presented the calculation of the response of the electrical tunnel
current and the current-induced spin transfer torque to ac and dc biases in
FM-I-FM tunnel junctions by means of the time-dependent nonequilibrium Green
function technique. In general, when a spin-polarized current is injected
into a ferromagnet, the polarized electrons may experience the fields such
as the external field, the anisotropy field, the dipolar-dipolar
interaction, and the demagnetization field\cite{heide1,miltat}, leading to
the precession of the polarized electrons around these fields. To conserve
the spin angular momentum, the magnetization also precesses around the
polarized direction of the injected electrons, giving rise to a spin
transfer which is quite different from a magnetic field induced by the
current. The ways to realize the switching of magnetization can be either by
a magnetic field, or by a magnetic field induced by the current, or by the
spin transfer effect\cite{bazaliy}. The way through the mechanism of spin
transfer offers a convenient and fast choice to switch magnetization, which
could be observed from other works and our present study.

In summary, we have investigated the spin-dependent tunneling in the
presence of an ac bias applied to a FM-I-FM system by considering the ac
tunneling current and the ac CISTT. We have formulated the time-averaged
current (time-averaged CISTT) in the form of a summation of dc current (dc
CISTT) multiplied by products of Bessel functions with the energy levels
shifted by $m\hbar \omega _{0}$. The tunneling current can be viewed as to
happen between the photonic sidebands of the two ferromagnets. Our
calculation shows that low-frequency ac field suppresses the current, and
the electrons may more easily tunnel through the barrier under a
high-frequency ac field as the response for a capacitance in a classical
case. It is found that the TMR almost does not vary with an ac bias, which
suggests that the ac electric field contributes less to the spin-dependent
scatterings. The current-induced spin transfer torque under an ac bias has
also been investigated. It has been shown that an ac bias may overall
suppress the spin transfer torque, but in a narrow regime of the ac bias,
the CISTT is greatly enhanced, characterized by a sharp peak which can be
viewed as the photon-assisted enhancement. It has been found that the
particular amplitude of the ac bias at which the CISTT shows a peak has a
linear relation with the frequency of the ac bias approximately. As a
consequence, people could adjust the proper values of the amplitude and the
frequency of the ac bias in accordance with such a linear relation to
realize the enhancement of the CISTT, which gives another possible option to
enhance the CISTT externally. In addition to using an dc current, our result
might give an alternative hint to control the local moments of ferromagnets,
and also to control the resistance states.

\section*{Acknowledgments}

This work is supported in part by the National Science Foundation of China
(Grant No. 90103023, 10104015), the State Key Project for Fundamental
Research in China, and by the Chinese Academy of Sciences.

\newpage

{\bf FIGURE CAPTIONS}

Fig. 1 Time dependence of the tunneling current, where the effective masses
of the left and the right ferromagnets are taken as unity, the molecular
fields are assumed to be $0.9eV$, the Fermi energy is taken as $1.295eV$
which are taken from Ref.\cite{moodera} for Fe, and $T_{1}=T_{4}=0.01$ eV, $%
T_{2}=T_{3}=0,$ the other parameters are assumed as $\hbar =1,$ $V_{0}=-0.1$
$V,$ $\omega _{0}=0.003$ eV, $\theta =\pi /3$ and the temperature is at $%
100K $.

Fig. 2 The frequency dependence of the time-averaged tunneling current. The
remaining parameters are assumed to be the same as those in Fig. 1.

Fig. 3 The time-averaged tunneling current as a function of the amplitude of
the ac bias, $V_{ac}.$ The remaining parameters are assumed to be the same
as those in Fig. 1.

Fig. 4 Time-dependent spin-transfer torque as a function of $V_{ac}$ under
different frequencies, where $t=15$ ($10^{-4}ns$), $V_{0}=-0.05V,$ and the
other parameters are assumed to be the same as those in Fig. 1.

Fig. 5 Time-dependent spin-transfer torque as a function of frequency $%
\omega _{0}$ under different $V_{ac}$, where $V_{0}=-0.1V,$ and the other
parameters are the same as those in Fig. 4.

Fig. 6 $V_{ac}$-dependence of the time-averaged spin transfer torque under
different frequencies $\omega _{0}$, where the other parameters are the same
as those in Fig. 1.

Fig. 7 Time-averaged spin transfer torque as a function of $\omega _{0}\ $%
for different $V_{ac}$, where the other parameters are the same as those in
Fig. 1.

Fig. 8 Time-averaged spin transfer torque versus $V_{ac}$ for different
frequencies when only the ac bias is present ($V_{0}=0$). The inset is the
particular amplitude, $V_{ac}^{p}$, at which the CISTT is peaked, versus
frequency $\omega _{0}$. The parameters are the same as those in Fig. 1.

\newpage

%%%%%%%%%%%%%%%%%%%%%%%%%%%%%%%%%%%%%%%%%%%%%%%%%%%%%%%%%%
\begin{figure}[htbp]
\centering
    \includegraphics[width = 9 cm]{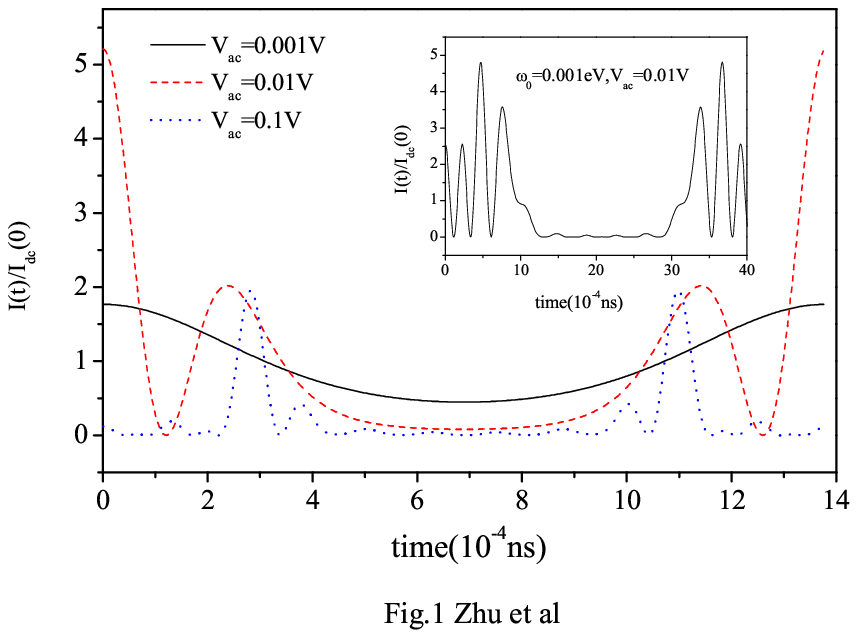}
    \caption{Time dependence of the tunneling current, where the effective masses
    of the left and the right ferromagnets are taken as unity, the
    molecular fields are assumed to be $0.9eV$, the Fermi energy is
    taken as $1.295eV$
    which are taken from Ref.\cite{moodera} for Fe, and $T_{1}=T_{4}=0.01$ eV, $%
    T_{2}=T_{3}=0,$ the other parameters are assumed as $\hbar =1,$
    $V_{0}=-0.1$
    $V,$ $\omega _{0}=0.003$ eV, $\theta =\pi /3$ and the temperature is at $%
    100K $.}
    \end{figure}
%%%%%%%%%%%%%%%%%%%%%%%%%%%%%%%%%%%%%%%%%%%%%%%%%%%%%%%%%%
%%%%%%%%%%%%%%%%%%%%%%%%%%%%%%%%%%%%%%%%%%%%%%%%%%%%%%%%%%
\begin{figure}[htbp]
\centering
    \includegraphics[width = 9 cm]{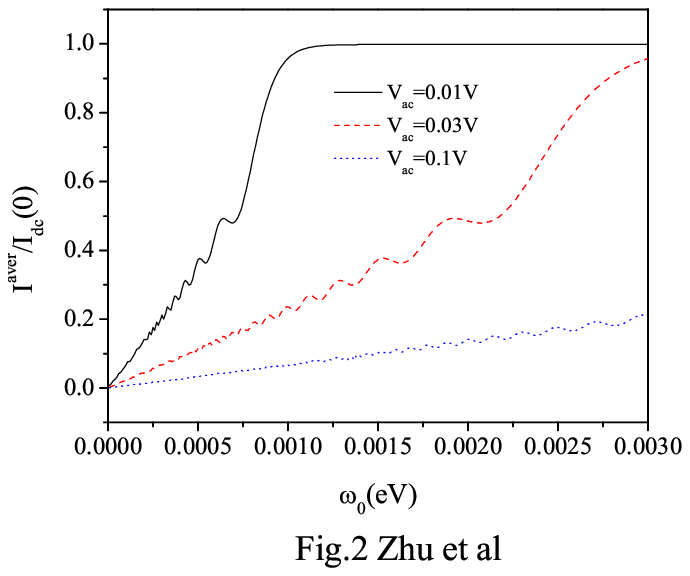}
    \caption{The frequency dependence of the time-averaged tunneling current. The
    remaining parameters are assumed to be the same as those in Fig.1.}
    \end{figure}
%%%%%%%%%%%%%%%%%%%%%%%%%%%%%%%%%%%%%%%%%%%%%%%%%%%%%%%%%%
%%%%%%%%%%%%%%%%%%%%%%%%%%%%%%%%%%%%%%%%%%%%%%%%%%%%%%%%%%
\begin{figure}[htbp]
\centering
    \includegraphics[width = 10 cm]{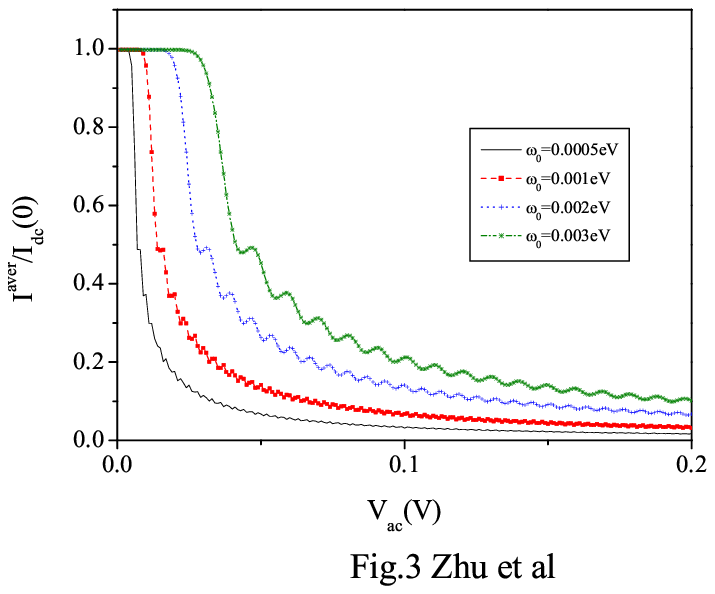}
    \caption{The time-averaged tunneling current as a function of the amplitude of
    the ac bias, $V_{ac}.$ The remaining parameters are assumed to be
    the same as those in Fig. 1.}
    \end{figure}
%%%%%%%%%%%%%%%%%%%%%%%%%%%%%%%%%%%%%%%%%%%%%%%%%%%%%%%%%%
%%%%%%%%%%%%%%%%%%%%%%%%%%%%%%%%%%%%%%%%%%%%%%%%%%%%%%%%%%
\begin{figure}[htbp]
\centering
    \includegraphics[width = 10 cm]{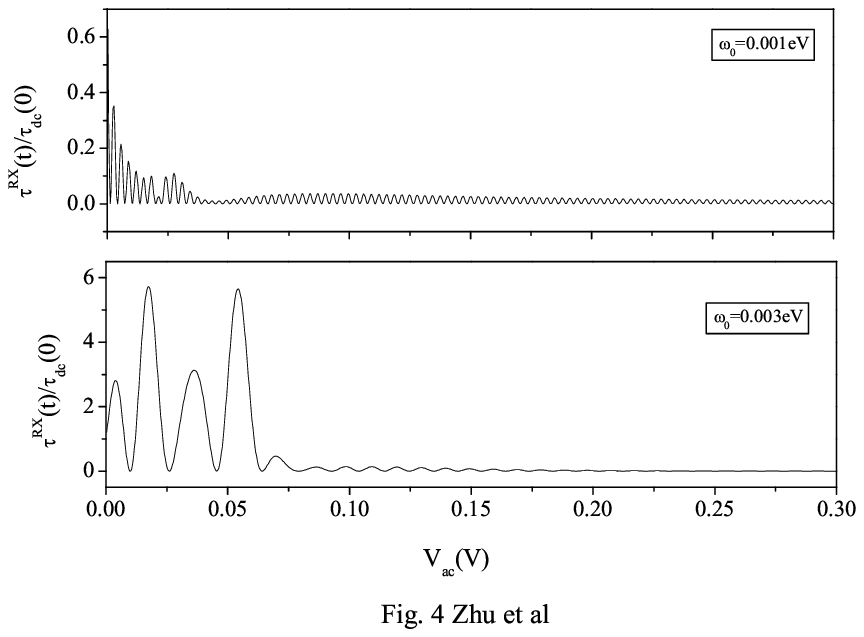}
    \caption{Time-dependent spin-transfer torque as a function of $V_{ac}$ under
     different frequencies, where $t=15$ ($10^{-4}ns$), $V_{0}=-0.05V,$
     and the other parameters are assumed to be the same as those in
     Fig. 1.}
    \end{figure}
%%%%%%%%%%%%%%%%%%%%%%%%%%%%%%%%%%%%%%%%%%%%%%%%%%%%%%%%%%
%%%%%%%%%%%%%%%%%%%%%%%%%%%%%%%%%%%%%%%%%%%%%%%%%%%%%%%%%%
\begin{figure}[htbp]
\centering
    \includegraphics[width = 10 cm]{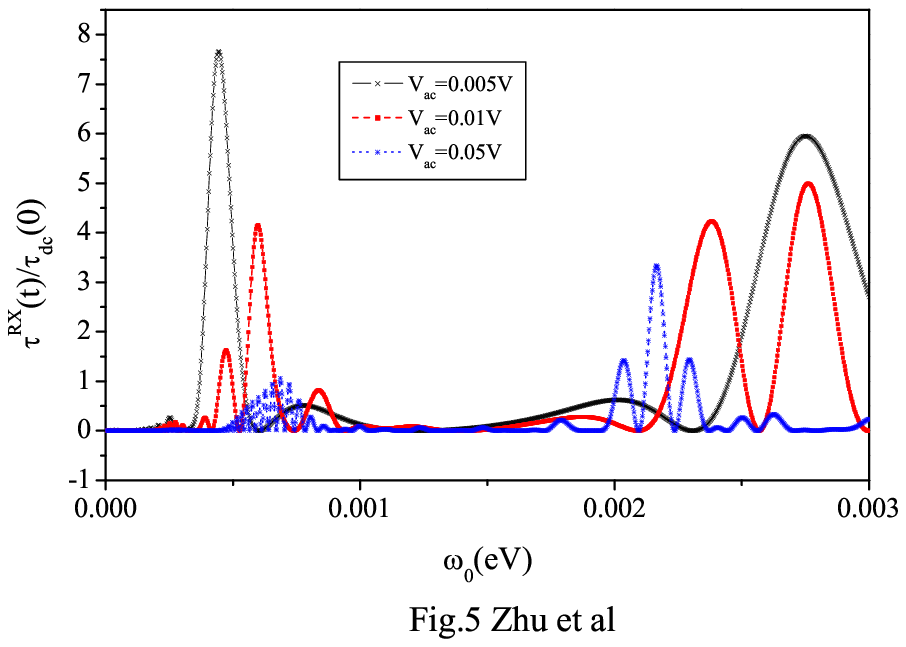}
    \caption{Time-dependent spin-transfer torque as a function of frequency $%
    \omega _{0}$ under different $V_{ac}$, where $V_{0}=-0.1V,$ and
     the other parameters are the same as those in Fig. 4.}
    \end{figure}
%%%%%%%%%%%%%%%%%%%%%%%%%%%%%%%%%%%%%%%%%%%%%%%%%%%%%%%%%%
%%%%%%%%%%%%%%%%%%%%%%%%%%%%%%%%%%%%%%%%%%%%%%%%%%%%%%%%%%
\begin{figure}[htbp]
\centering
    \includegraphics[width = 10 cm]{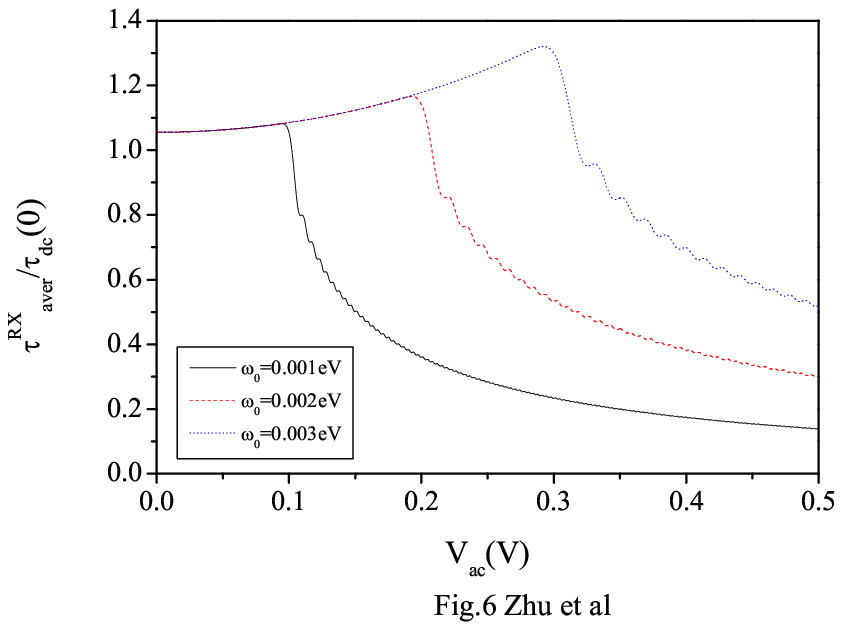}
    \caption{$V_{ac}$-dependence of the time-averaged spin transfer torque under
    different frequencies $\omega _{0}$, where the other parameters
    are the same as those in Fig. 1.}
    \end{figure}
%%%%%%%%%%%%%%%%%%%%%%%%%%%%%%%%%%%%%%%%%%%%%%%%%%%%%%%%%%
%%%%%%%%%%%%%%%%%%%%%%%%%%%%%%%%%%%%%%%%%%%%%%%%%%%%%%%%%%
\begin{figure}[htbp]
\centering
    \includegraphics[width = 10cm]{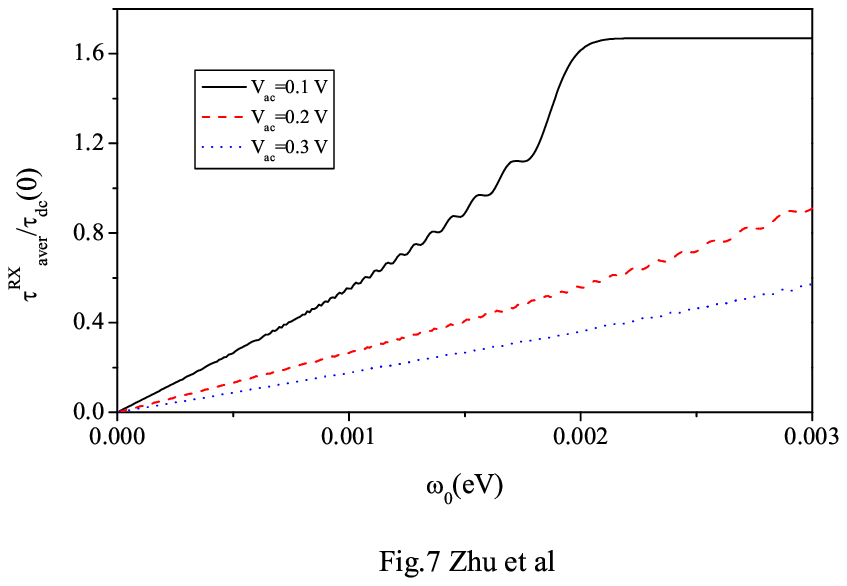}
    \caption{Time-averaged spin transfer torque as a function of $\omega _{0}\ $%
    for different $V_{ac}$, where the other parameters are the same as
    those in Fig. 1.}
    \end{figure}
%%%%%%%%%%%%%%%%%%%%%%%%%%%%%%%%%%%%%%%%%%%%%%%%%%%%%%%%%%
%%%%%%%%%%%%%%%%%%%%%%%%%%%%%%%%%%%%%%%%%%%%%%%%%%%%%%%%%%
\begin{figure}[htbp]
\centering
    \includegraphics[width = 10 cm]{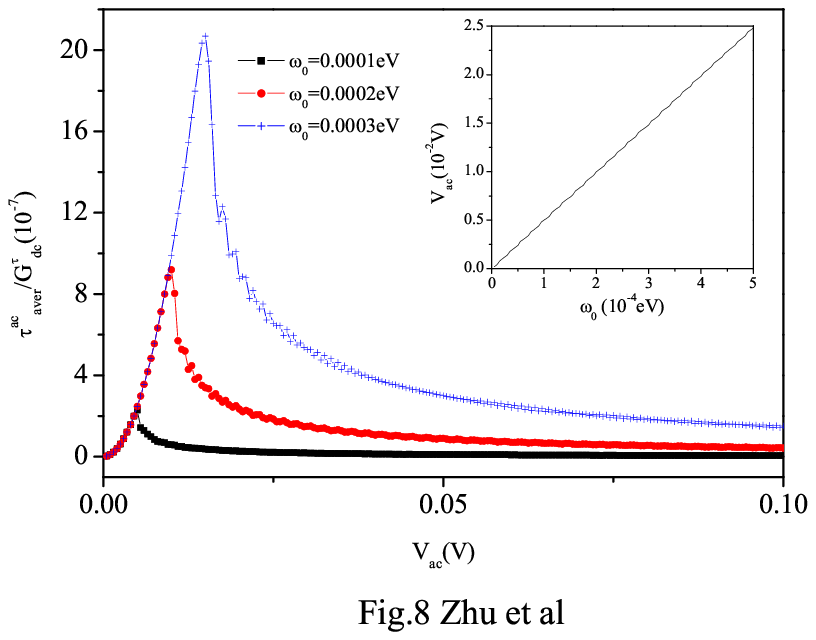}
    \caption{Time-averaged spin transfer torque versus $V_{ac}$ for different
    frequencies when only the ac bias is present ($V_{0}=0$). The
    inset is the particular amplitude, $V_{ac}^{p}$, at which the
    CISTT is peaked, versus frequency $\omega _{0}$. The parameters
    are the same as those in Fig. 1.}
    \end{figure}
%%%%%%%%%%%%%%%%%%%%%%%%%%%%%%%%%%%%%%%%%%%%%%%%%%%%%%%%%%


\begin{thebibliography}{1}

\bibitem[\mbox{}]{Permanent address.} $^{\ast }$Corresponding author.
E-mail: gsu@@gscas.ac.cn.

\bibitem{wolf} S. A. Wolf, et al, Science {\bf 294,} 1488 (2001); J. S.
Moodera, J. Nassar and G. Mathon, Annu. Rev. Sci. {\bf 29,} 381 (1999); G.
A. Prinz, Science {\bf 282, }1660 (1998).

\bibitem{heide} C. Heide, A. I. Krikunov, Yu. F. Ogrin, P. E. Zilberman, and
R. J. Elliott, J. Appl. Phys. {\bf 87}, 5221 (2000).

\bibitem{slonczewski} J. C. Slonczewski, J. Magn. Magn. Mater. {\bf 159}, L1
(1996).

\bibitem{berger} L. Berger, J. Appl. Phys. {\bf 55,} 1954 (1984); Phys. Rev.
B {\bf 54,} 9353 (1996).

\bibitem{tsoi} M. Tsoi et al, Phys. Rev. Lett. {\bf 80,} 4281 (1998).

\bibitem{wegrowe} J. -E. Wegrowe et al., Europhys. Lett. {\bf 45, }626
(1999).

\bibitem{sun} J. Z. Sun, J. Magn. Magn. Mater. {\bf 202,} 157 (1999).

\bibitem{myers} E. B. Myers et al, Science {\bf 285}, 867 (1999).

\bibitem{asamitsu} A. Asamitsu, Y. Tomioka, H. Kuwahara, and Y.Tokura,
Nature {\bf 388,} 50 (1997).

\bibitem{katine} J. A. Katine, F. J. Albert, R. A. Buhrman, E. B. Myers, and
D. C. Ralph, Phys. Rev. Lett. {\bf 84,} 3149 (2000); J. Grollier, et al.,
Appl. Phys. Lett. {\bf 78,} 3663 (2001); J. Z. Sun, D. J. Monsma, D. W.
Abraham, M. J. Rooks, and R. H. Koch, Appl. Phys. Lett. {\bf 81,} 2202
(2002).

\bibitem{myers1} E. B. Myers, F. J. Albert, J. C. Sankey, E. Bonet, R. A.
Buhrman, and D. C. Ralph, Phys. Rev. Lett. {\bf 89,} 196801 (2002).

\bibitem{s2} J. C. Slonczewski, Phys. Rev. B {\bf 39,} 6995 (1989).

\bibitem{zhang} Ya. B. Bazaliy, B. A. Jones, and Shou-Cheng Zhang, Phys.
Rev. B {\bf 57,} R3213 (1998).

\bibitem{fnf} K. B. Hathaway and J. R. Cullen, J. Magn. Magn. Mater. {\bf %
104-107,} 1840 (1992); A. Brataas, Yu. V. Nazarov, and G. E. W. Bauer, Eur.
Phys. J. B {\bf 22,} 99 (2001); M. D. Stiles, and A. Zangwill, J. Appl.
Phys. {\bf 91, }6812 (2002).

\bibitem{waintal} X. Waintal, E. B. Myers, P. W. Brouwer, and D. C. Ralph,
Phys. Rev. B {\bf 62, }12317 (2000).

\bibitem{fsf} Y. Tserkovnyak and A. Brataas, Phys. Rev. B {\bf 65, }094517
(2002).

\bibitem{waintal1} X. Waintal and P. W. Brouwer, Phys. Rev. B {\bf 63, }%
220407 (2001).

\bibitem{waintal2} X. Waintal and P. W. Brouwer, Phys. Rev. B {\bf 65, }%
054407 (2002) .

\bibitem{heide1} C. Heide, P. E. Zilberman, and R. J. Elliott, Phys. Rev. B
{\bf 63,} 064424 (2001); C. Heide, Phys. Rev. Lett. {\bf 87, }197201 (2001).

\bibitem{szhang} S. Zhang, P. M. Levy, and A. Fert, Phys. Rev. Lett. {\bf %
88, }236601 (2002).

\bibitem{szhang1} \ Z. Li, and S. Zhang, cond-mat/0302337; \
cond-mat/0302339.

\bibitem{zhu1} For a detailed discussion, see Zhen-Gang Zhu, Gang Su, Biao
Jin, and Qing-Rong Zheng, Phys. Lett.A {\bf 306}, 249 (2003).

\bibitem{su} Gang Su and M. Suzuki, Mod. Phys. Lett. B {\bf 16}, 711 (2002).

\bibitem{haug} H. Haug, and A. -P. Jauho,{\footnotesize \ }{\it Quantum
Kinetics in Transport and Optics of Semiconductors} (Springer-Verlag,
Berlin, 1998).

\bibitem{jauho} A. P. Jauho, N. S. Wingreen, and Y. Meir, Phys. Rev. B {\bf %
50}, 5528 (1994).

\bibitem{anantram} M. P. Anantram, and S. Datta, Phys. Rev. B {\bf 51}, 7632
(1995).

\bibitem{bwang} Baigeng Wang, Jian Wang, and Hong Guo, Phys. Rev. Lett. {\bf %
82, }398 (1999).

\bibitem{note1} As the calculations here are quite involved, for the purpose
of illustration we may assume that the elements of the tunneling matrix do
not depend on the momentum for simplicity, which is also a reasonable
assumption in a wide-band limit. The $k$ and $q$ dependence of the tunneling
matrix will be left for future study.

\bibitem{tkng} Tai-Kai Ng, Phys. Rev. Lett. {\bf 76, }487 (1996).

\bibitem{zhu} Zhen-Gang Zhu, Gang Su, Qing-Rong Zheng, and Biao Jin, Phys.
Lett. A {\bf 300,} 658 (2002).

\bibitem{Tucker} J. R. Tucker, and M. J. Felman, Rev. Mod. Phys. {\bf 57, }%
1055 (1985).

\bibitem{jauho1} A. P. Jauho, cond-mat/9711141.

\bibitem{moodera} J. S. Moodera, M. E. Taylor, And R. Meservey, Phys. Rev. B
{\bf 40, }11980 (1989).

\bibitem{miltat} J. Miltat, et al., J. Appl. Phys. {\bf 89}, 6982 (2001).

\bibitem{bazaliy} Ya. B. Bazaliy, B. A. Jones, and S. C. Zhang, J. Appl.
Phys. {\bf 89,} 6793 (2001).

\end{thebibliography}
\end{document}